\documentclass{PoS}

\title{First results of the two square meters multilayer glass composite mirror design proposed for the Cherenkov Telescope Array developed at INFN}

\ShortTitle{INFN multilayer glass composite mirror design proposed for CTA\,\hspace{3.8cm}\normalfont\normalsize Schultz et al.}

\author{C. Schultz$^{a}$, M. Doro$^{a,b}$, L. Lessio$^{a,c}$, M. Mariotti$^{a}$, R. Rando\footnote{Speaker}\,\,$^{a}$, for the CTA Consortium\footnote{Full consortium author list at http://cta-observatory.org}\\

{\footnotesize
$^{a}$ Universit\`a di Padova and INFN, I-35131 Padova, Italy\\
$^{b}$ Max-Planck-Institut f\"ur Physik, D-80805 M\"unchen, Germany\\
$^{c}$ INAF, I-35122 Padova, Italy}\\
E-mail: \email{cornelia.schultz@pd.infn.it}
}

\abstract{The Cherenkov Telescope Array (CTA) is a future ground-based gamma-ray astronomy detector that will consist of more than 100 Imaging Atmospheric Cherenkov Telescopes of different sizes. The total reflective surface of roughly 10 000 m$^2$ requires unprecedented 
technological efforts towards a cost-efficient production of light-weight and reliable mirror substrates at high production rate. 
We report on a new mirror concept proposed for CTA developed by INFN, which is based on the replication from a spherical convex mold under low pressure. 
The mirror substrate is an open structure design made by thin glass layers at the mirror's front and rear interspaced by steel cylinders.
A first series of nominal size mirrors has been produced, for which we discuss the optical properties in terms of radius of curvature and 
focusing power.}

\FullConference{The 34th International Cosmic Ray Conference,\\
		30 July- 6 August, 2015\\
		The Hague, The Netherlands}

\begin{document}

\section{Introduction}
An Imaging Atmospheric Cherenkov Telescope (IACT) is a detector that is suited for the study of the very high energy (VHE) gamma-ray component of the cosmic radiation. IACTs are designed to detect Cherenkov light with a pixelized camera placed in the focal plane of a large reflector. Such light is emitted in the Earth's atmosphere by electromagnetic showers, which are initiated by the interaction of gamma rays with the atmospheric molecules. 

The future generation of IACTs is represented by the Cherenkov Telescope Array (CTA) project~\cite{actis11,acharya13} that aims at a further improved performance with respect to the current generation of such instruments like MAGIC, H.E.S.S. and VERITAS. To achieve such progress in terms of enhanced sensitivity over an extended range in the VHE band and beyond, the CTA design foresees the installation of IACTs of different sizes, sensitive to different energies of the VHE band. In total, CTA will consist of more than 100 telescopes comprising several tens of small size telescopes (SSTs) of 6 m diameter, few tens of medium size telescopes (MSTs) of 12 m diameter and few large size telescopes (LSTs,~\cite{cortina15}) of 23\,m diameter that will be split across two sites - one in the Northern and one in the Southern hemisphere. Each telescope requires a dish tessellated with many light-weight\footnote{For the LST, the specification is $<50$\,kg.}, robust and reliable mirror facets of adequate reflectance ($>85$\% into the focal spot from 300 to 550\,nm) and  focusing quality (the point spread function (PSF) in terms of D$_{80}$, i.e. the diameter of 80\% light containment, but demanding very little maintenance. In the case of the LST the D$_{80}$ of the individual mirror facets is required to be less than 1/3 of the pixel diameter at the focal length of the telescope, which is 16.6\,mm (0.59\,mrad for $f=28$\,m) in the case of a pixel diameter of 2 inch ($\sim$50\,mm) including the light concentrator~\cite{cortina15}. Usually, IACTs are not protected by domes meaning that the mirrors are continuously exposed to the environment. The challenge for CTA is to develop low-cost, mirrors of 1 to 2 m$^2$ area, hexagonal geometry, high optical performance in the wavelength range of interest and long-term durability, with the potential for a high production rate. The technologies currently under investigation pursue different methods involving different mirror layouts and materials in order to produce mirrors that meet the CTA specifications~\cite{foerster13a,pareschi13}. Besides conventional layouts based on the so-called Davies-Cotton single-mirror configuration, which are adapted for the MST and the LST, some alternative designs based on dual-mirror layouts that adopt the so-called Schwarzschild-Couder configuration for the MST and the SST are investigated. For such dual-mirror designs, the requirements on the primary and secondary mirror are more constraining than for the MST and LST demanding short radii of curvature and high asphericity elements~\cite{pareschi13}.

\section{Production technique under development}

In collaboration with the Compositex Srl\footnote{\href{http:www.compositex.com}{http:www.compositex.com}} and LT-Ultra\footnote{\href{http:www.ltultra.com}{http:www.ltultra.com}} companies, INFN Padova contributed significantly to the production of the reflective surfaces of the two 17 m diameter MAGIC telescopes, MAGIC I and MAGIC II, for which two different types of composite mirrors were produced~\cite{doro08,pareschi08}. To contribute to the LST, INFN Padova is developing a new process to produce mirrors of hexagonal shape of 1.5~m flat-to-flat (FTF) diameter based on the production techniques used for the MAGIC mirrors. Besides this mirror design, an alternative mirror layout developed by ICRR, Tokyo, Japan in cooperation with the Sanko company~\cite{hayashida15} is proposed for the LST.

\begin{figure}[!h]
\center
    \includegraphics[width=0.95\textwidth]{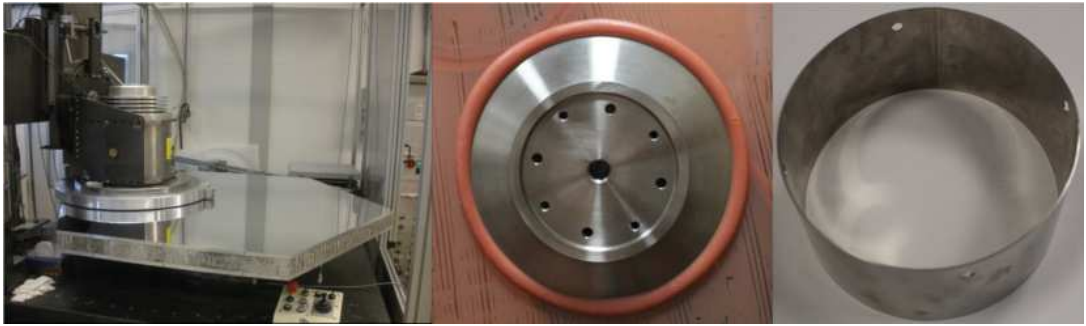}
  \caption{Left: Diamond-milled 1.5 FTF diameter aluminum mold. Center: Pad for the active mirror control attachment. Right: Image of a stainless steel cylinder.}
  \label{fig:icrc2015-204-01}
 \end{figure}

The multilayer glass composite mirror has an open structure design and consists of a thin rear and front glass multilayer interspaced by steel cylinders. In such way, the mirror is rigid and light-weight at the same time. The technique is based on the curvature replication under low pressure from a convex diamond-milled aluminum mold (Figure~\ref{fig:icrc2015-204-01} left panel), with a nominal radius of curvature of $(57.9\pm 0.5)$\,m a microroughness of $<30$\,nm. 

To ensure a good surface contact between the mold and the glass sheets, small channels have been incised into the mold after its polishing, allowing the evacuation of entrapped air via a vacuum pump that is attached through a central hole creating a pressure of $(0.8\pm0.04)$\,bar. 

\begin{figure}[!h]
\center
    \includegraphics[width=0.35\textwidth]{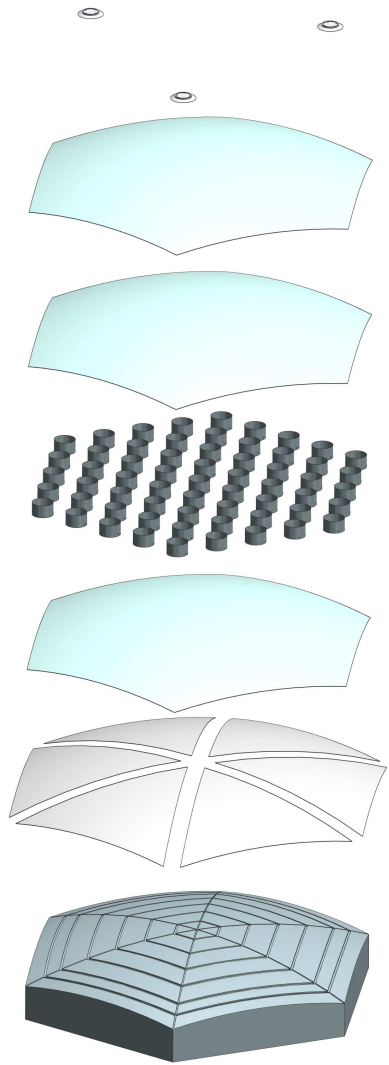}
  \caption{Production scheme of the multilayer glass composite mirror based on the replication from a spherical convex mold.}
  \label{fig:icrc2015-204-02}
 \end{figure}

From the back to the reflecting surface (top to bottom in Figure~\ref{fig:icrc2015-204-02}):
\begin{itemize}
\item 3 120~mm diameter stainless steel pads for the connection of the actuators of the active mirror control (AMC,~\cite{cortina15})
\item 1.8~mm thick borosilicate\footnote{\href{http:www.eot.it}{http:www.eot.it}} glass of hexagonal shape   
\item 1.8~mm thick borosilicate glass of hexagonal shape
\item 63 120~mm diameter stainless steel cylinders of 60~mm height and 0.5~mm thickness
\item 1.8~mm thick borosilicate glass of hexagonal shape
\item 6 equilateral triangles with 866~mm side length of 2.0~mm thick Schott\footnote{\href{http:www.schott.com}{http:www.schott.com}} BOROFLOAT\textsuperscript{\textregistered} 33 Al + SiO$_2$ +  HfO$_2$ + SiO$_2$ front coated~\cite{foerster13b} glass   
\end{itemize}

In a first step, the mirror rear glass multilayer is assembled, i.e. the glass sheets are spherically shaped on the mold under low pressure and are glued together, curing the glue, a two component epoxy paste, at room temperature. Contemporaneously, the pads (Figure~\ref{fig:icrc2015-204-01}) for the AMC are glued to the back of the mirror rear at a circle of 130\,cm with respect to the mirror center. Similarly, the mirror front is assembled on the mold with the difference that the reflective surface is composed by equilateral triangles produced by the BTE\footnote{\href{http:www.bte.de}{http:www.bte.de}} company whose reflective surfaces, an Al+3-layer coating with a  reflectance $>85$\% in the wavelength of interest~\cite{foerster13b}, are protected by a residual free removable foil. Our decision to assemble the reflective surface from smaller elements is mainly driven by economic reasons. After curing the mirror rear, the cylinders (Figure~\ref{fig:icrc2015-204-02}) are glued to the concave side. 4 holes of 5~mm diameter are drilled to the cylinders to allow entrapped water to drain off. Once cured, the rear is glued to the convex side of the mirror front glass multilayer while curved under low pressure on the mold.

\section{Results}

\begin{figure}[!h]
\center
    \includegraphics[width=0.95\textwidth]{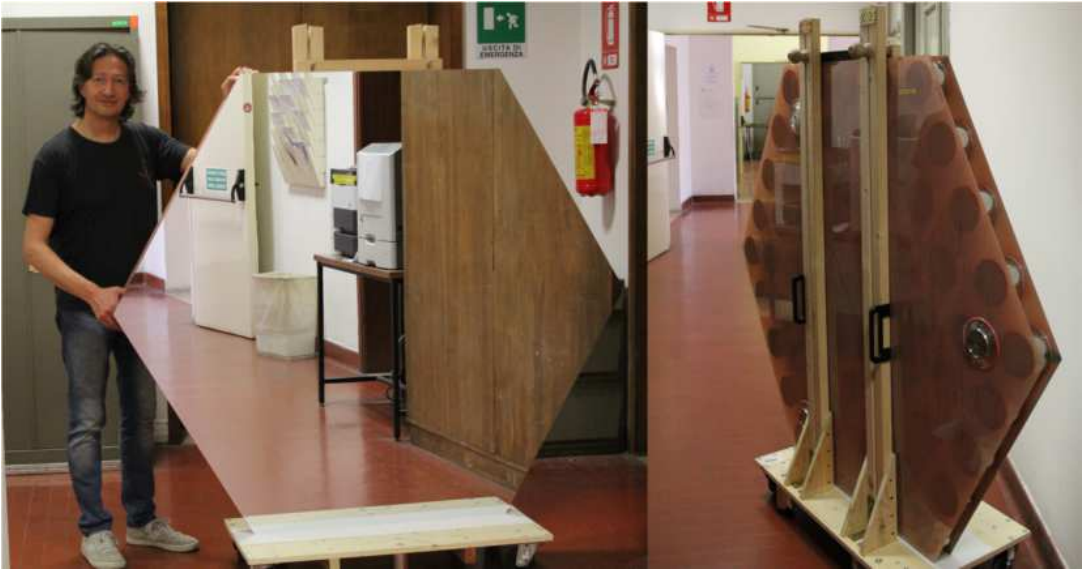}
  \caption{Front (left) and rear (right) of one of the first 1.5 FTF diameter mirror prototypes proposed to be used for the LST.}
  \label{fig:icrc2015-204-03}
 \end{figure}

At the beginning of this year, several prototypes (Figure~\ref{fig:icrc2015-204-03}) of a weight of roughly 42~kg have been produced and measurements of the radius of curvature and the PSF have been performed with a so-called $2f$-setup~\cite{foerster13a}, by determining the distance, where the spot reflected by the mirror, which is illuminated by a point-like source, is minimum and performing an analysis of the spot image taken with a CCD camera. 

\begin{figure}[!h]
\center
    \includegraphics[width=\textwidth]{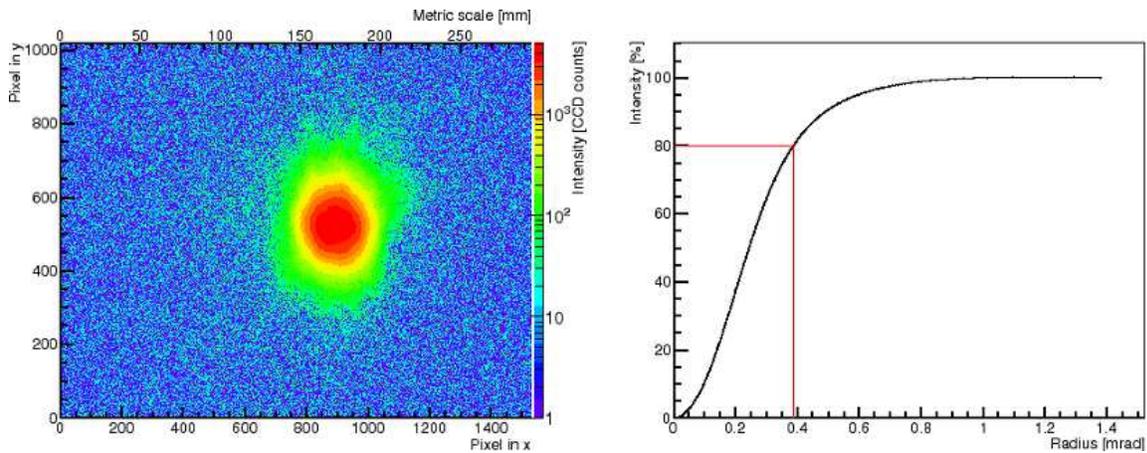}
  \caption{Left: CCD image of the minimum focal spot at the distance of 58.3 m. On top a metric scale is drawn. Right: Focal spot profile. The red vertical line indicates the R$_{80}$.}
  \label{fig:icrc2015-204-04}
 \end{figure}

We found the radius of curvature to be consistent with the nominal radius of curvature ($57.9\pm 0.5$\,m) of the mold with variations up to $\sim$+40~cm ($\sim$0.7\%), while the PSF expressed as the D$_{80}$ was calculated to be as good as $\sim$0.8\,mrad (Figure \ref{fig:icrc2015-204-04}). We note that the light source used for the measurement is slightly extended ($\sim$1.5\,mm in diameter) which needs to be taken into account when evaluating the PSF. Thus the actual PSF is $\sim$0.75\,mrad.
 
The deviations of the measured radius of curvature with respect to that of the mold are rather small and are no issue for a parabolic optical design as foreseen for the LST in which different radii of curvature are required~\cite{pareschi13,hayashida15}. The PSF is most likely affected by local variations from the mold geometry due to dust and due to the surface waviness of the reflective surface glass leading to variations of $\sim$+0.16\,mrad ($\sim$27\%) with respect to the CTA specifications of the LST ($<$0.59 mrad). 

\pagebreak
\section{Conclusions and outlook}

We are finalizing the development of an open structure composite layout for mirrors suited for IACTs based on the shape replication from a mold. The design consists of thin rear and front glass multilayers interspaced by steel cylinders ensuring the mirror to be rigid and light-weight. For the LST, part of the AMC system is designed to be integrated in one of the mirror corners~\cite{hayashida15}, which still has to be implemented in the mirror design presented here. The mirror prototypes produced show a good agreement with the nominal radius of curvature, with variations of few tens of centimeters. The PSF achieved so far is slightly higher than the CTA specifications for the LST, which is mostly limited by the surface waviness of the reflective surface glass and dust on the mold. A solution to improve the PSF could be to use a front glass of higher quality for the reflective surface and potentially to use different segmentation geometries. We are already planning the construction of few more prototypes with such improvements.

We plan to test the durability of some of the prototypes produced at a dedicated outdoor facility. Mechanical testing and a cross check of the optical characteristics of our prototypes are foreseen to be carried out at adequate facilities accessible to the CTA consortium~\cite{foerster13a}. These facilities are also equipped to measure the reflectance into the focal spot.

\acknowledgments{We gratefully acknowledge support from the agencies and organizations 
listed under Funding Agencies at this website: \href{http://www.cta-observatory.org/}{http://www.cta-observatory.org/}.}

\end{document}